\documentclass[aps,pre,twocolumn,nofootinbib,showpacs,preprintnumbers,amsmath,amssymb,superscriptaddress]{revtex4} 

\usepackage{graphicx}
\usepackage{bm}
\usepackage{psfrag}
\usepackage{subfigure}
\usepackage{color}

\newcommand{\be}{\begin{equation}}
\newcommand{\ee}{\end{equation}}
\newcommand{\ba}{\begin{eqnarray}}
\newcommand{\ea}{\end{eqnarray}}

\begin{document} 

\author{Carolina Brito}
\affiliation{Service de Physique de l'\'Etat
Condens\'e,~CEA-Saclay; URA 2464, CNRS, F-91191~Gif-sur-Yvette,~France}
\affiliation{Instituut-Lorentz, LION,
Leiden University, P.O. Box 9506, 2300 RA Leiden, Netherlands}
\author{Olivier Dauchot}
\affiliation{Service de Physique de l'\'Etat
Condens\'e,~CEA-Saclay; URA 2464, CNRS, ~91191~Gif-sur-Yvette,~France}
\author{Giulio Biroli}
\affiliation{Institut de Physique Th\'eorique, CEA-Saclay, URA 2306, CNRS F-91191
Gif-sur-Yvette, France}
\author{Jean-Philippe Bouchaud}
\affiliation{Science and  Finance, Capital Fund Management, 6 Bd
Haussmann, 75009 Paris, France}

\date{\today}

\title{Elementary Excitation Modes in a Granular Glass above Jamming}

\begin{abstract}
The dynamics of granular media in the jammed, glassy region is described in terms of ``modes'', by applying a Principal Component Analysis (PCA) to the covariance matrix of the position of individual grains. We first demonstrate that this description is justified and gives sensible results in a regime of time/densities such that a metastable state can be observed on long enough timescale to define the reference configuration. For small enough times/system sizes, or at high enough packing fractions, the spectral properties of the covariance matrix reveals large, collective fluctuation modes that cannot be explained by a Random Matrix benchmark where these correlations are discarded. We then present a first attempt to find a link between the softest modes of the covariance matrix during a certain ``quiet'' time interval and the spatial structure of the rearrangement event that ends this quiet period. The motion during these cracks is indeed well explained by the soft modes of the dynamics before the crack, but the number of cracks preceded by a ``quiet'' period strongly reduces when the system unjams, questioning the relevance of a description in terms of modes close to the jamming transition, at least for frictional grains.
\end{abstract}


\maketitle

\section{Introduction}

The physical processes by which super-cooled liquids, granular systems, and colloids acquire rigidity are not well understood. At first sight, the phenomenon of rigidity is utterly trivial: we know that when we move one end of a ruler the other end moves the same distance. 
It is so simple that, as P.W. Anderson noticed, {\it it is hard to realize that such an action at a distance is not built into the laws of nature except in the case of the long-range forces such as gravity and electrostatics...We are so accustomed to this rigidity property that we don't accept its almost miraculous nature, that is an ``emergent property" not contained in the simple law of physics, although it is a consequence of them}. 

Recently, intense research has been devoted to this problem and it has become clear that the emergence of rigidity in soft matter is likely to be related to a collective phenomenon. Many hints came from numerical and analytical studies of the jamming transition of hard and elastic frictionless spheres \cite{PhysRevE.68.011306, vanHecke_JPCM2010}. In this case, it has been shown that when the system acquires rigidity it has no redundant mechanical constraints. As a consequence, it is in a marginally stable, isostatic, state. This has dramatic consequences for the vibrational spectrum, which displays a broad band of soft modes \cite{wyart_EPL2005, PhysRevE.72.051306}. The role of these modes in the dynamics close to the rigidity transition and, more generally, for glassy liquids has been emphasized in \cite{Brito_JCP2009, PhysRevLett.102.038001, PhysRevLett.97.258001}. However, the applications and verifications of these theoretical ideas in experiments are scarce. A first attempt has been performed for colloidal glasses~\cite{bonn}, but the limitation in experimental resolution does not allow to draw definitive conclusions. Here we focus on mechanically driven granular media. These are the physical systems that triggered the studies of anomalous properties of vibrational modes and isostatic properties \cite{PhysRevB.48.15646,PhysRevE.70.061302}. Despite of this, there is still no experimental study in the literature on the role of the modes close to the rigidity transition. The aim of our work is to present a first analysis of the modes close to the rigidity transition of vibrated frictional grain assemblies. Note that the presence of friction is expected to modify the properties of the transition compared to the ideal case of hard spheres \cite{PhysRevE.75.020301, Henkes2010}. In particular, our system seems to be characterized by micro-cracks of all scales, leading to `jumps' in the position of particles with a power-law distribution of sizes~\cite{JP-Levyflights}, which makes the analysis in terms of modes particularly tricky. Still, we believe that the tools we developed are interesting also from a methodological point of view, and will be useful for analysing other systems that undergo a jamming transition. 

In \cite{fred1} it was shown that as the packing fraction of a horizontally vibrated monolayer of bidisperse hard grains is increased beyond a certain packing fraction $\phi_J$, the system is able to support mechanical stresses. This is the rigidity transition, which appears as a genuine critical point, where a dynamical correlation length and a correlation time simultaneously diverge, showing that the dynamics occurs by involving progressively more collective rearrangements $\phi_J$. Contrary to the case of frictionless hard sphere or colloids the pressure does not diverge at $\phi_J$ but at a higher density. 

Experimentally, we have access to the covariance matrix of the positions, ${\bf C}_p$~\footnote{We have also studied the covariance matrix of the instantaneous velocities but at the present stage of the study, it did not provide further insight. We thus concentrate here on the results given by the study of ${\bf C}_p$}. Whether and to what extent this can be interpreted in terms of vibrations along some modes is one of the main open questions that we shall address. We shall also investigate how the eigenstates and eigenvalues of ${\bf C}_p$ evolve when approaching $\phi_J$ and their relation with the  dynamics. In order to do that, we have to separate signal from noise in the eigenproperties of ${\bf C}_p$. This is a common and crucial problem in dealing with covariance matrices, which will be addressed by using tools and concepts previously developed and used in other fields like finance and biology \cite{PhysRevLett.83.1467, micheletti, martinez, PhysRevLett.91.198104, PhysRevLett.103.268101, bouchaud2009far}.

\section{Experimental system and preliminaries on the particle positions covariance matrix ${\bf C}_p$}
\label{exp_system}
The experimental set-up and the quench protocols are described in detail in~\cite{fred1}. A 1:1 bidisperse monolayer of 8500 brass cylinders of diameters $d_{s} = 4 \pm 0.01 mm$ and $d_b = 5\pm0.01 mm$ stands on a glass plate which is horizontally vibrated at a frequency of $10$ Hz and an amplitude of $10 mm$. The grains are confined within a fixed rectangular metal frame of width L $\approx 100$ $d_s$. 
The packing fraction $\phi$ can be adjusted by moving a lateral wall on which we control the pressure. 
The stroboscopic motion of a set of 1500 grains in the center of the sample is tracked with an accuracy of $2.10^{-3} d_s$. Lengths are measured in $d_s$ units and time in cycle units. The initial protocol produces a very dense state with a packing fraction of $\phi = 0.8457$. The packing fraction is then decreased by very small steps down to $0.84$. For each $\phi$, the plate vibrates $10^4$ cycles during which the pressure at the wall is stored. 
At high packing fraction, the mean pressure is dominated by the static pressure, which is measured by interrupting the vibration. At some $\phi$, the kinetic part of the pressure becomes dominant and this is identified as the jamming transition, which takes place at $\phi_J \in [0.8417, 0.8422]$. 

The main properties of the grain displacements have been discussed in detail in~\cite{fred1,fred2}. Very recently, we re-examined these statistics of the displacements and found the rather surprising results alluded to above, which we report in another paper of the present special issue~\cite{JP-Levyflights}. 
First, let us insist on the fact that the typical displacement of the particles is of the order of one hundredth of its diameter. Accordingly, all structural rearrangements are frozen on the experimental timescales: the neighbors of a given particle do not change during the experimental time scale. 
Second, the motion of the particles, sub-diffusive at short times and diffusive at asymptotically large times, exhibits a super-diffusive motion at intermediate timescales close to the jamming transition. Our recent analysis of the data shows that this superdiffusion is not induced by long-range temporal correlations of the velocity field, as we first surmised in~\cite{fred1}. Quite on the contrary, the displacements on the intermediate timescale are made of a large number of incoherent jumps with a broad distribution of jump sizes. 
However, these jumps become more and more collective as the systems becomes rigid at $\phi_J$, which appears as a genuine critical point, where the dynamical correlation length diverges.

As stated in the introduction, our aim here is to further characterize the dynamics and its spatial organization close to the rigidity transition by studying the covariance matrix of the particles positions, defined as:
\be
{\bf C}_p = \langle~ \delta r_{i,\alpha}~\delta r_{j,\beta}~\rangle_T=\langle~  (r_{i,\alpha}-\langle r_{i,\alpha}\rangle_T)~(r_{j,\beta}-\langle r_{j,\beta}\rangle_T)~\rangle_T,\nonumber
\ee
where $r_{i,\alpha}$ is the $\alpha=x$ or $y$ Cartesian coordinate of the $i^{th}$ grain and $\langle~.~\rangle_T$ denotes the temporal average over an observation window of duration $T$.

For solids at thermal equilibrium, the modes of ${\bf C}_p$ can be identified with structural vibrational modes because particles simply oscillate around their equilibrium positions. For example for crystals at low enough temperature the matrix ${\bf C}_p$ is equal to the temperature times the inverse of the Hessian matrix of the potential energy evaluated for the ground state configuration. In this case the eigenvectors of ${\bf C}_p$ are plane waves that identify with the phonons.  In the present case, the system being driven out of equilibrium, it is not warranted at all that the modes of ${\bf C}_p$ can be interpreted as vibrational modes. However, as argued in the introduction and as will be confirmed in the following, studying the spectral properties of ${\bf C}_p$ remains a powerful tool of investigation, {\it provided that} the particles have a well defined average position: ${\bf C}_p$ measures the fluctuations around a metastable state and its spectral properties allow one to interpret these fluctuations in terms of effective excitation modes. 

\begin{figure*}[t!]
\begin{center}
 \rotatebox{-90}{\resizebox{12cm}{17.0cm}{\includegraphics{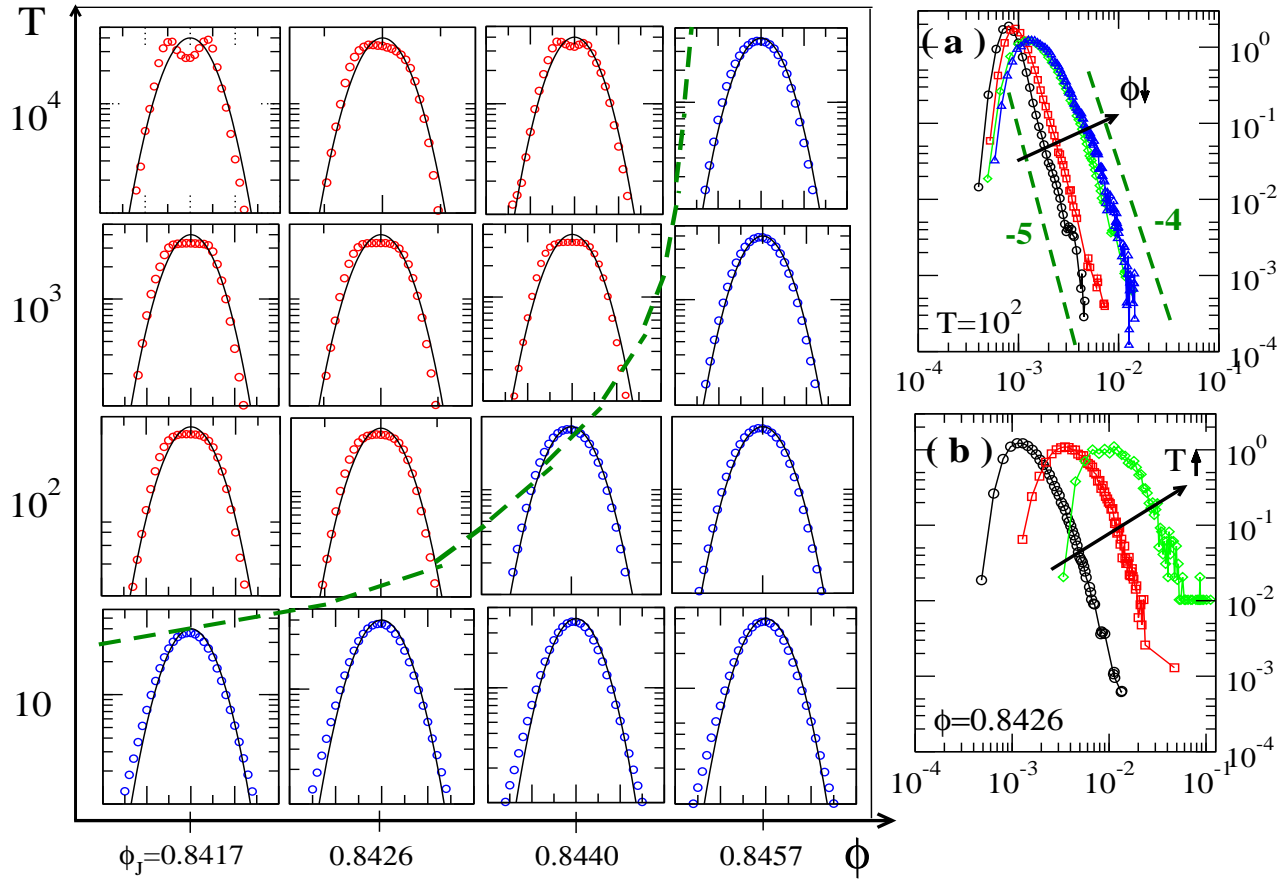}}}
\caption{Left: Distributions of the position fluctuations $\rho(\delta x_i/\sigma_i)$ for four values of the packing fraction and four durations of observation $T$.  We focus on the top of the distribution, which is compared to a Gaussian (continuous line).
Right: {\bf(a)} $\rho(\sigma_i )$ for different $\phi$ keeping $T=100$ constant. The arrow indicates the direction of decreasing $\phi \in [0.8417, 0.8426, 0.8440, 0.8457]$
{\bf(b)} $\rho(\sigma_i )$ for different $T$ keeping $\phi=0.8426$ constant. The arrow indicates the direction of increasing $T \in [10^2, 10^3, 10^4]$.}
\vspace{-0mm}
\label{diagram}
\end{center}
\end{figure*}

We thus start by investigating the fluctuations of the particle positions around the average position. Here we focus on a single component $x$ of the position, but we have checked that the conclusions are identical for both, confirming that the dynamics is isotropic as already observed in~\cite{fred1}. For a given particle $i$, one can compute the average position $\langle x_i \rangle_T$ on a time $T$ and the fluctuations around it $\delta x_i(t)= x_i(t)- \langle x_i \rangle_T$. Note that in glassy disordered systems, this average position can itself evolve with time and be an extra source of fluctuations. The variance of $\delta x_i$ over time $T$ characterizes how far the particle is, typically, from its average position:  $\sigma_i^2 = \langle \delta x_i^2\rangle_T$. We find (see below) that $\sigma_i$ significantly fluctuates from particle to particle, reflecting the presence of dynamical heterogeneities in the system: while some particles hardly move during time $T$, others are able to ``rattle'' quite a bit (but still on scales much smaller than the grain diameter!). More precisely, the distributions $\rho(\sigma_i)$ are shown on the right of Fig.~\ref{diagram}. When decreasing the packing fraction towards $\phi_J$, $\rho(\sigma_i)$ shifts to larger values of $\sigma_i$, indicating larger overall motions for each particle, as expected.  As $\phi$ decreases, $\rho(\sigma_i)$ also broadens significantly demonstrating more and more heterogeneities among the particles. Indeed describing the right tail of the distribution by a power law: $\rho(\sigma_i)\sim \sigma_i^{-1-\mu}$, one find $\mu$ decreasing from $\approx 4$ to $\approx 3$, when decreasing $\phi$ towards $\phi_J$. As a matter of fact, for the largest packing fractions, the power-law tail is so steep that it can as accurately be described by an exponential. When $T$ increases, the distribution $\rho(\sigma_i)$ shifts to larger values of $\sigma_i$ as expected, but does not broaden, indicating that the heterogeneities are already well developed within an interval of time $T=100$. Such observations are yet another confirmation of the statistical properties of the dynamics studied in~\cite{fred1,JP-Levyflights}. Note that the exponent $\mu$ here should not be confused with the exponent describing the tail of individual jump sizes, as defined in~\cite{JP-Levyflights}: here, we characterize the variation of the vibrations across different grains, and not for a single grain over time. In order to perceive the difference more clearly, imagine a case where all particles perform {\it exactly} the same motion, be it a regular random walk or a L\'evy flight: in both cases, $\rho(\sigma_i)$ should then be a delta function since there is no dispersion at all.

We then study the distribution of rescaled positions, $\delta x_i/\sigma_i$, by averaging over all times and all particles. The distributions are computed for four different packing fractions $\phi$ and four durations $T$ of the window of observation. They are then ensemble averaged over the $10^4/T$ intervals provided by the full dataset. From now on all statistical quantities (such as the eigenvalue 
spectra, etc.) are evaluated this way, without further specifying it except when necessary to avoid confusion.

The distributions shown in Fig.~\ref{diagram} highlight some important characteristics of the dynamics. The parameter space $(\phi,T)$ can be divided into two regions, as illustrated by the hatched line: for small enough observation duration $T$ or large enough packing fractions, the distributions are unimodal with a Gaussian core: particles jiggle around a well defined average position; for longer $T$ or smaller densities, the distribution starts developing a flat top, with a poorly defined maximum. This suggest that on these longer observation times, the average position of a significant part of the particles is not well defined anymore. Particles either drift slowly or even find (collectively) another metastable position, as suggested by the double peak observed in the case $\phi=0.8417$ and $T=10^4$, i.e. for the loosest packing fraction and the longest observation time. This means that over long time scales, the evolution of the average position becomes comparable or even larger than the fluctuations, and it becomes meaningless to describe the system in terms of small vibrations around a fixed metastable state. For an infinite size system, some rearrangement always happens somewhere, and the covariance matrix ${\bf C}_p$ is always ill-defined. The ``allowed'' time scale $T_{\max}(\phi,L)$ is expected to scale inversely with the system size; however, when $T_{\max}$ becomes too small, statistical noise becomes dominant and prevents a reliable estimation of the spectrum of ${\bf C}_p$.
In the following sections, we will navigate between these constraints and try to identify well defined eigenmodes of the motion. 

\section{Spectral properties of ${\bf C}_p$}

In this section we shall study in detail the spectral properties of ${\bf C}_p$. Our aim is twofold: first, as stated above, the spectral properties are affected by measurement noise for finite $T$. Thus it is important to disentangle trivial properties of ${\bf C}_p$ induced by the noise from relevant ones, which we do in the first subsection. Second, we would like to understand whether ${\bf C}_p$ is indeed measuring some steady fluctuations around a well-defined metastable state. We will refer to such a property as robustness and study it in the next subsection. Third, we will see that the structure of the modes itself confirms that the $10$ first modes are significantly out of the noise range. In order to do so we concentrate on one specific case in the middle range of our parameter space, $\phi=0.844$ and $T=100$, for which particle positions seem to be  well defined on the observation window duration and we consider the whole set of tracked particles $N=1500$. Given that the correlation matrix is computed in a observation window $T < 2N$, there are at the best only $T$ non-zero eigenvalues amongst $2N$. The eignevalues are normalized by $\bar{\sigma}^2/Q$, where $\bar{\sigma}=\langle \sigma_i\rangle_i$ is the average of the $\sigma_i$'s over all particles and $Q=T/2N$ is the total number of measured data points $2 N \times T$ divided by the total number of variables $4N^2$. With such a normalization, one can easily compare the spectra of ${\bf C}_p$ for systems with different average mobility $\bar{\sigma}$ as well as for computations of ${\bf C}_p$ with different values of $Q$ -- for instance when considering subsystems of smaller size $N$, as we shall do in the next section. 

\subsection{The role of noise}

In order to obtain some hints on the role of noise in the spectral properties of ${\bf C}_p$ we will compare our results to the ones obtained by constructing the covariance matrix with iid random variables $z_i(t)=\eta_i(t) \sigma_i$, where $\eta_i(t)$ are iid Gaussian variables and $\sigma_i$ are positive random variables following the {\it experimental} distribution $\rho(\sigma_i)$. Note that $\sigma_i$ is constant during each interval of duration $T$. In this benchmark model, which we shall refer to as the Random Matrix case ($RM_{\sigma}$), the spatial correlations between $\eta_i(t)$ and $\eta_j(t)$, $j \neq i$, and between the different $\sigma$'s are discarded. By comparison, we will be able to evaluate the relevance of these correlations in the experimental system. Our results will also be compared to the case of a 2-d equilibrium crystal.   

\begin{figure}[t!]
\begin{center}
\vspace{-5mm}
\rotatebox{-90}{\resizebox{!}{8.0cm}{\includegraphics{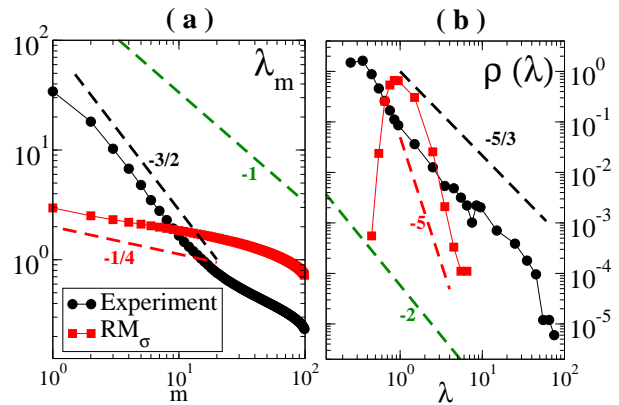}}}
\caption{Spectral properties for the whole system ($\phi=0.844$, $N=1550$ particles, $T=100$). {\bf (a)}: Normalized spectrum  $\lambda_m$ vs. $m$ for the experimental data and for the Random Matrix case ($RM_{\sigma}$). The (blue) dotted line with exponent $-1/\alpha=-1$ is the prediction for the 2D Crystal. {\bf (b)}: The associated density of states. Dashed lines are guide for the eyes.}
\vspace{-5mm}
\label{fullsystem}
\end{center}
\end{figure}

Figure~\ref{fullsystem} displays the normalized spectrum $\lambda_m$ and the associated density of states $\rho(\lambda)$ for both the experimental data and the $RM_\sigma$ simulation when $\phi=0.844$ and $T=100$.
Note that there is a straightforward correspondence between the behaviour of the more commonly studied $\rho(\lambda)$ at large $\lambda$ and that of $\lambda_m$ at small $m$ since $\frac{1}{N} m(\lambda)$ is exactly the inverse cumulated distribution of $\lambda$. Accordingly a powerlaw behaviour $\lambda_m \sim m^{-1/\alpha}$ translates into a density $\rho(\lambda)\sim \lambda^{-(1+\alpha)}$. 
For instance a 2D crystal, with a density of states $\rho(\omega)\sim \omega^{d-1}=\omega$, with $\omega \sim 1/\lambda^{1/2}$, has $\rho(\lambda)\sim \lambda^{-2}$, that is $\alpha=1$ and $\lambda_m \sim m^{-1}$ as indicated on the figure by the blue dotted line.
For the Random Matrix case ($RM_{\sigma}$), if the distribution $\rho(\sigma_i)$ has power-law tails with exponent $\mu$, then the top eigenvalues of the correlation matrix also has a power-law tailed distribution, with exponent $1+\mu/2$ and $\rho(\lambda)$ should  decay at large $\lambda$ at least as slow as $\lambda^{-(1+\mu/2)}$. In the present case, $\mu \approx 4$ and one would expect $\alpha \approx 2$, whereas we measure here $1/\alpha \approx 1/4$. The reason for this discrepancy is insufficient sampling: as is clear from Fig.~\ref{diagram}, right, there is hardly a factor $10$ contrast between the largest $\sigma$ encountered in the sample and the typical one as given by the median of the distribution. We have check that with many more samples, the expected power-law tail 
eventually appears. But we have been careful here to take exactly the same statistics in the simulation and in the experiment, so that the comparison made in Fig.~\ref{fullsystem} is meaningful.
The ten largest eigenvalues for the  experimental system are therefore clearly larger than for the Random Matrix case. Our data is consistent with $\alpha=2/3$, that is a slower decay of the spectrum than for both the $RM_{\sigma}$ and the crystal case. This comparison shows unambiguously that the top eigenvalues of ${\bf C}_p$ contain useful information about the dynamics of the system, and are {\it not} drowned in noise.
It also demonstrates the existence of strong spatial correlations: by moving together, particles achieve large collective fluctuations that would not develop otherwise.

\subsection{Micro-cracks and robustness}

As already stressed, in such an heterogeneous system the above analysis strongly relies on the selection of the observation windows, in order to ensure that the system remains in a single metastable state. We thus compute the instantaneous self density-correlation function: 
\be
C_q(t,t_0) = \langle \cos(\vec q \cdot [\vec r_i(t) - \vec r_i(t_0)] )\rangle_i,
\ee
where $\vec r_i(t)$ is the particles position at time $t$ and $\vec q$ is a wave vector whose amplitude is given by $q=\pi/a$. $a$ is chosen as a small length scale of the order of $a^{*}=\langle(\vec r_i(t+\tau^*) - \vec r_i(t))^2\rangle^{1/2}=7.\, 10^{-3}$, where $\tau^*$ is the timescale at which dynamical heterogeneities are maximal (see~\cite{JP-Levyflights} in the present volume for details). The average is computed over all particles, but not on the initial time $t_0$ and it is not ensemble averaged either. $C_q(t,t_0)$ decays to zero when the average displacement of the particles between time $t_0$ and $t_0+t$ is larger than $a$. 

\begin{figure}[t!]
  \begin{center}
    \rotatebox{-0}{\resizebox{!}{8.0cm}{\includegraphics{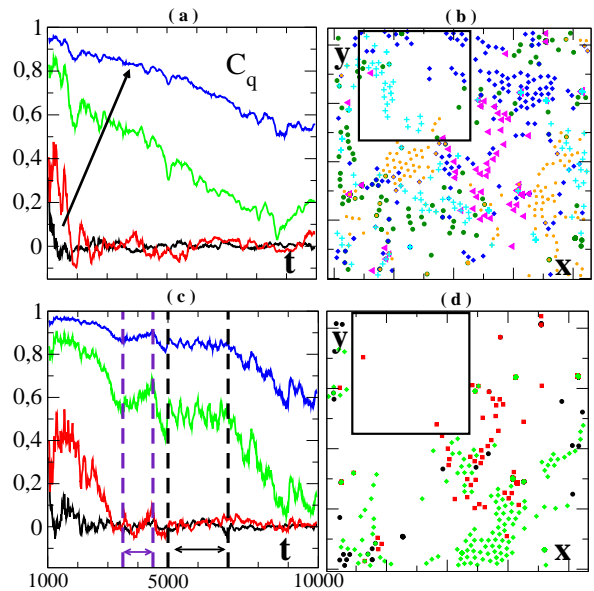}}}
    \caption{Relaxation events for $\phi=0.844$.Top: {\bf(a)}: Instantaneous self density-correlation function $C_q(t,t_0)$ as defined in the text. Each curve is for a different $a \in [a^*, 2a^*, 5a^*, 10 a^*]$ and the arrow indicates increasing $a$. {\bf(b)} positions of the particles which where identified as significantly contributing to the relaxation of the system (see definition in the text). Different symbols (and colors online) indicate different time windows of duration $T=1000$. The square box surrounds the rather inactive region, which we select as the subset of particles over which we re-compute $C_q(t,t_0)$ displaid in {\bf(c)}. The dotted lines indicate the temporal windows isolated as quiet periods. {\bf(d)}: Same plot as in (b) but restricted to the period of time $t \in [3000 -- 6000]$.}
\vspace{-5mm}
\label{Cq_data7}
\end{center}
\end{figure}

Ideally, one would like to observe sudden drops of $C_q(t,0)$ that signal moments when a significant collective event occurs, hopefully separated by long enough ``quiet'' periods. Also, the same plateaus and cracks should be present for a reasonable range of length-scales $a$.  This is not the case here, as clearly observed in Fig.~\ref{Cq_data7}(a). For all $a \in [a^*, 10a^*]$, the decrease of $C_q(t,0)$ is progressive rather than taking place during sudden drops. Furthermore, the observation time windows which look like quiet for a given $a$, are in fact very jerky when decreasing $a$. The reason for these features are (i) the heterogeneity of the relaxation (for a large enough system, some relaxation event is always taking place somewhere in the system) and (ii) the scale invariance of these relaxation events, or ``micro-cracks'' as recently pointed out in~\cite{JP-Levyflights}.

This scale invariance makes it very hard (if not impossible) to define properly the metastable states of the system, and the corresponding covariance matrix ${\bf C}_p$. It suggests to identify relaxation events not by their size but through an iterative process such as the one proposed in~\cite{PhysRevLett.102.088001} to identify cage jumps in the trajectories of particles in a super-cooled liquid. In a nutshell the algorithm consists in cutting each trajectory in two sub-trajectories, in such a way that each subset maximizes a clustering criteria, and in applying iteratively the algorithm to each subset until the maximization criterion is no more significant. As a result one obtains for each particle a set of instants corresponding to the times when it has relaxed, {\it without} specifying the amplitude required to relax. Fig.~\ref{Cq_data7}(b) displays such events: at all times some small regions of the system relax, each of them contributing to a small decrease of $C_q(t,0)$. 

Altogether, when approaching the jamming transition from above, the system as a whole becomes more and more heterogeneous, less rigid, and metastable states harder and harder to define. This makes the computation of $\bf{C}_p$ increasingly difficult, precisely where we would like to use it to characterize the dynamics. 
However, one also notices in Fig.~\ref{Cq_data7}(b) a region indicated by the square box, where there is little activity as compared to the rest of the system. Fig.~\ref{Cq_data7}(c) again displays $C_q(t,0)$ but averaged on the particles belonging to this quiet region only. One now can better identify periods of time where $C_q(t,0)$ is rather constant, independently of $a$. These sub-regions are rigid during long enough time intervals to perform the analysis in terms of modes. In the following we shall refer to these sub-systems and time-interval as the ``rigid subsets'' of the system. 

\begin{figure}[t!]
\begin{center}
\rotatebox{-90}{\resizebox{!}{8cm}{\includegraphics{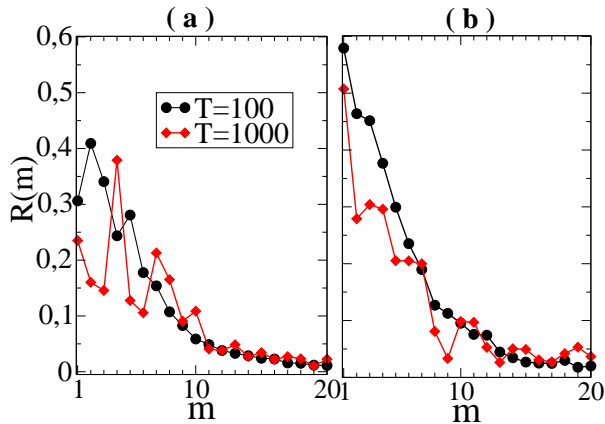}}}
\caption{Robustness of the modes for $\phi=0.844$ as characterized by $R(m)$ for $T=100$ and $T=1000$; {\bf(a)}: whole set of particles, {\bf(b)} subset of particles and observation window identified in fig.~\ref{Cq_data7} ($N=350$ particles, $T=100$ and $T=1000$).}
\vspace{-5mm}
\label{Rob_data7}
\end{center}
\end{figure}

This we confirm by assessing the robustness of the modes. For that purpose, we compute the following indicator:
\be
R(m) = \sum_{j=-M}^{j=+M} \langle \lambda_m |  \lambda'_{m+j} \rangle^2, 
\ee
where $\langle \lambda_m |  \lambda'_{m +j} \rangle$ is the scalar product between the modes computed during two successive observation windows of duration $T$. If the two eigenbases are precisely the same, $R(m)$ is equal to 1 for all eigenvectors $\lambda_m$. Note that the definition allows that the modes computed in one observation window project onto any of the (2M+1) modes of the second basis surrounding mode $m$, in order to allow the neighbouring modes to possibly exchange their rank. For $M \ge 2$ and not too large the results are basically independent of $M$. Here we fixed $M=2$. One observes in Fig.~\ref{Rob_data7} that the robustness of the modes is twice larger when restricting the analysis to the rigid subsets. 

\begin{figure}[t!]
\begin{center}
\rotatebox{-90}{\resizebox{!}{8cm}{\includegraphics{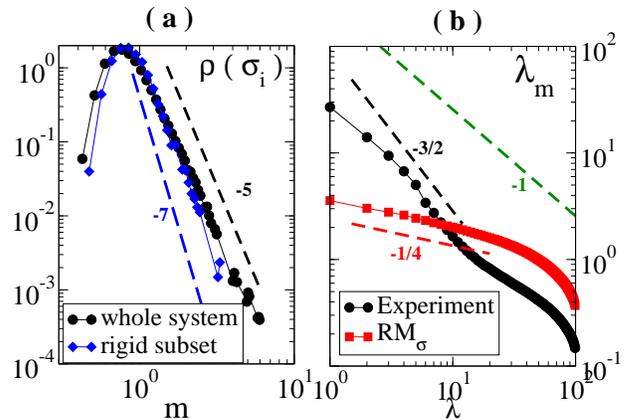}}}
\caption{Spectral properties computed for the subset of particles and observation window identified in fig.~\ref{Cq_data7} ($\phi=0.844$, $N=350$ particles, $T=100$). {\bf (a)}: Density of $\sigma_i$ for both the whole system and the subset of interest {\bf (b)}: Normalized spectrum  $\lambda_m$ vs. $m$ for the experimental data and for the Random Matrix case ($RM_{\sigma}$). The (blue) dotted line with exponent $-1/\alpha=-1$ is the prediction for the 2D Crystal. Dashed lines are guide for the eyes.}
\vspace{-5mm}
\label{Spect_data7_subset}
\end{center}
\end{figure}

Let us finally describe the spectral properties within the rigid subsets. Figure(\ref{Spect_data7_subset}) displays the distribution $\rho(\sigma_i)$ and the spectrum $\lambda_m$, which we compare to the ones obtained for the whole system (fig.\ref{diagram}-right and \ref{fullsystem}-a). We shall come back to the description of the distributions $\rho(\lambda)$ in the next section. The right tail of the distribution $\rho(\sigma_i)$ is more narrow ($\mu \simeq 6$) for the subset than for the whole system ($\mu \simeq 4 $) confirming that these rigid subsets are more homogeneous. For the $RM_\sigma$ case, the spectrum $\lambda_m$ for small $m$ is again flatter than expected, due to insufficient sampling ($1/\alpha \simeq 1/4$ instead of $1/3$). 
More remarkable is the fact that the spectrum for the experimental system remains well above the $RM_\sigma$ case and that it is almost identical to the one obtained in the whole system, suggesting that (i) it is not dominated by the shape of the distribution $\rho(\sigma_i)$ but on the contrary unveils non trivial correlations; (ii) the heterogeneities associated to these correlations are present at all scales.

Altogether, despite rather poor statistics and a significant amount of noise in the spectral properties of ${\bf C}_p$, the difference reported between the random matrix and the experimental cases confirms that one can trust the largest eigenvalues and that the spectrum in the experimental system is mostly governed (in its top region) by non-trivial spatial correlations.

\subsection{The structure of top eigenvectors}

We can substantiate this last assertion even more by comparing the localization properties of the associated eigenvectors. We start with the $RM_\sigma$ case.  Assuming that the $\sigma_i$ are power law distributed with an exponent $\mu$, the maximum $\sigma_{max}$ is given by the equation $N\int_0^{\sigma_{max}}\rho(\sigma_i)d\sigma_i\simeq O(1)$, which leads to $\sigma_{max}\propto N^{1/\mu}$.
Calling $i^*$ the value of $i$ corresponding to the maximum $\sigma_i$, one expects in the absence of correlations that the covariance matrix has a very large diagonal entry at $i^*$ (in the present case, the two eignevalues corresponding to the $x$ and $y$ directions are equally large as imposed by the equally large $\sigma_i$ in both directions). A reasonable guess, that can be justified using arguments as the one developed in \cite{GBP}, is that this leads to the largest eigenvalue and that the rest of the covariance matrix can be considered as a perturbation. Accordingly the largest eigenvalue in the $RM_\sigma$ is given by $\sigma_m^2$ and the corresponding eigenvector is completely localized on $i^*$. We have checked that both facts are indeed very well realized. Note that similar results hold for the second, third,.. largest eigenvalues which are related to the second, third largest value of $\sigma_i$.

\begin{figure}[t!]
  \begin{center}
   \rotatebox{-90}{\resizebox{!}{8cm}{\includegraphics{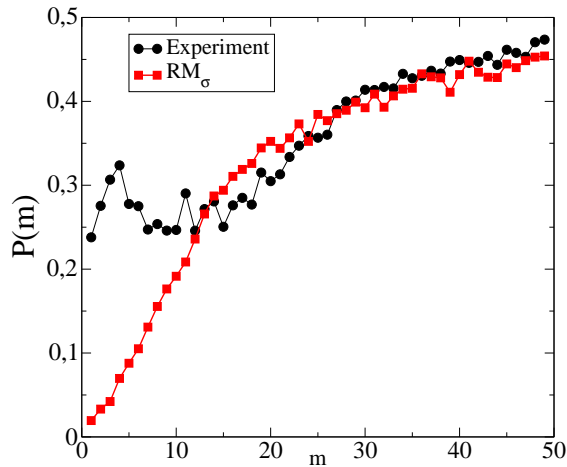}}}
    \caption{Participation ratio of the modes computed within a rigid subset for both the experimental data  and the Random Matrix case ($RM_{\sigma}$); $\phi=0.844$.}
    \label{pr_data7}
  \end{center}
\end{figure}

When considering the experimental covariance matrix, the largest eigenvalues and their corresponding eigenvectors are instead very different. This allows us to make clear that they are not due at all to noise and that spatial correlations are very instrumental in creating large eigenvalues or large fluctuations in the particle positions. In order to quantify this effect, we compute the participation ratio defined as:
\be
P(m) = \frac{1}{N \sum_{i=1}^{N} |\vec u_m^i|^4},
\ee
where $\vec u_m^i$ is the normalized displacement of particle $i$ within mode $m$. This quantity is such that, if the mode $m$ is completely localised on one particle, $P(m)= 1/N$. The other extreme case is when all the particles contribute equally to the mode: in this case $P(m) = 1$. Fig.~\ref{pr_data7} displays $P(m)$ for both the experimental and the Random Matrix cases. It is clear that largest eigenvalues for the $RM_\sigma$ case have a very small participation ratio, as expected since they are essentially localized on one or a few sites. Instead, the modes from experiments are characterized by a much higher $P(m)$, indicating that these modes are delocalized although less than plane waves for which $P(m)=2/3$. Beyond $m=10$, the participation ratio in both cases are very similar, suggesting that these bulk modes are incoherent and dominated by the local fluctuations of $\sigma_i$, and not by spatial correlations.

After this long but necessary description of the methodology, we can venture into the investigation of the relevant eigenvalues and modes structure, when approaching the jamming transition. On the basis of the above analysis, we will concentrate the computation of ${\bf C}_p$ on the rigid subsets and restrict the analysis to, say, the $10$ largest eigenvalues and corresponding eigenvectors.

\section{Towards the jamming transition : mode structure and dynamics}
\label{tjt}

In the following, we present a quantitative analysis of the modes and their properties when approaching the rigidity transition. This is to our knowledge the first attempt of this kind for granular assemblies. We first characterize the mode structure and then assess the role of these modes in the dynamical evolution of the system. 

We focus on the spectral properties of ${\bf C}_p$ for the four densities  $\phi \in [0.8417, 0.8426, 0.8440,  0.8457]$  following the methodology outlined in the previous section, i.e. identifying rigid subsets for which we measure ${\bf C}_p$. Then for a given density, we average over all the available rigid subsets. For the two densest cases, we identified two sub-regions which are rigid during typically $3000$ cycles. Closer to the jamming transition, there is no region, which remains rigid during more than $400$ cycles. We identified 6 of such rigid subsets for  $\phi= 0.8426$ and 5 for $\phi_J=0.8417$. In all cases the regions have about $N=350$ particles. An important observation is that it becomes increasingly difficult to measure the modes when approaching the rigidity transition. This is likely related to the findings explained in the companion paper \cite{JP-Levyflights} which show that at $\phi_J$ the dynamics is due to temporally incoherent but spatially correlated Levy jumps, corresponding to micro-cracks of all amplitudes that span the system, making it hard to find sub-regions where nicely separated, ``big'' cracks occur.

\begin{figure}[t!]
\begin{center}
\rotatebox{-90}{\resizebox{!}{8cm}{\includegraphics{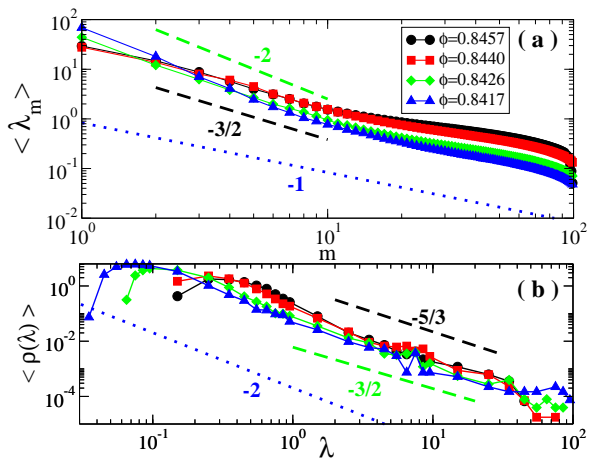}}}
\caption{{\bf (a)} Normalized spectrum $\lambda_m$ vs. $\times m$ for $\phi \in [0.8457, 0.8440, 0.8426, 0.8417]$. As specified in the text, ${\bf C}_p$ is computed within the subset of particles ($N\simeq350$), with well identified rigid periods. {\bf (b)} Corresponding eigenvalue densities $\rho(\lambda_m)$.} 
\label{lambda_dif_phi}
\end{center}
\end{figure}

\begin{figure}[t!]
\begin{center}
\rotatebox{-90}{\resizebox{!}{8.5cm}{\includegraphics{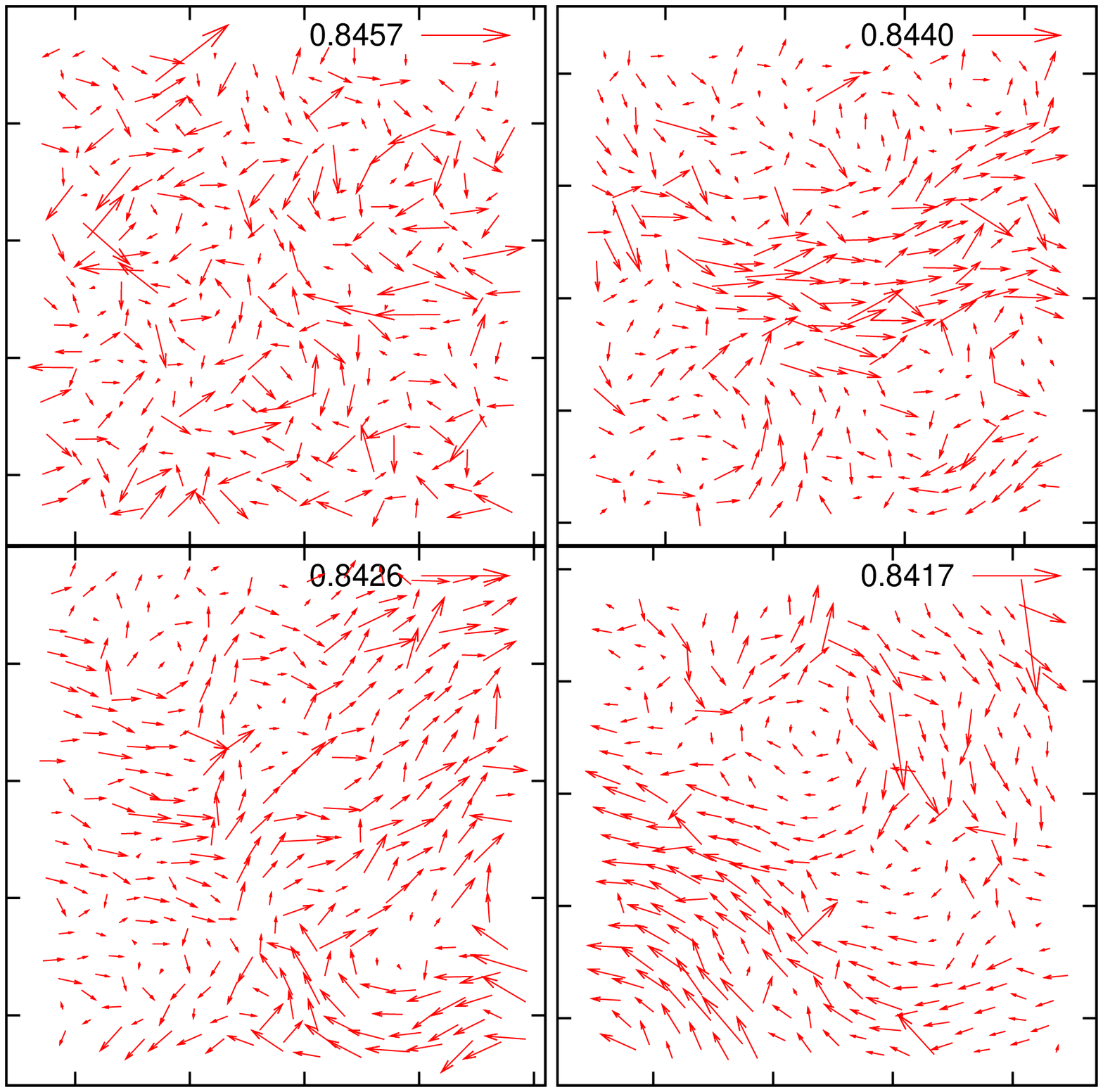}}}
\rotatebox{-90}{\resizebox{!}{8cm}{\includegraphics{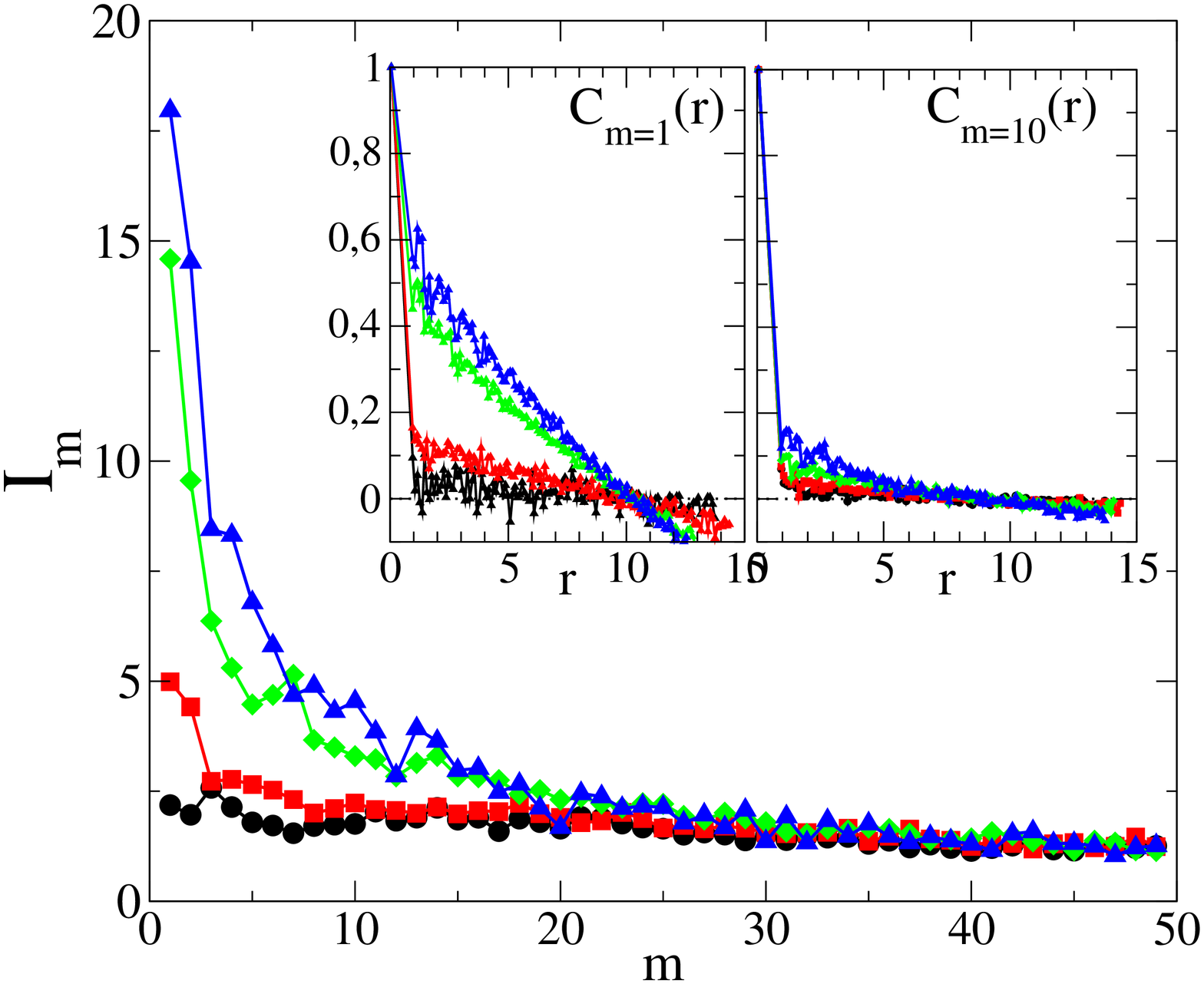}}}
\caption{Top: Four realisations of the first mode ($m=1$) for each packing fraction $\phi \in [0.8457, 0.8440, 0.8426, 0.8417]$. Bottom: Spatial correlations within the modes. The main plot provides an estimation of the correlation length as a function of the mode rank for the four packing fractions. Insets: spatial correlation for the four packing fractions for the modes $m=1$ and $m=10$.}
\label{modes_dif_phi}
\end{center}
\end{figure}

\subsection{Structure of the modes close to $\phi_J$}

The study of the spectral properties when approaching $\phi_J$ unveils that the softest modes become both softer and more extended:
\begin{itemize}
\item {\it Mode softness}:  As observed on figure~\ref{lambda_dif_phi}, the largest eigenvalue increases when approaching $\phi_J$; moreover $\lambda_m$  vs. $m$ becomes steeper in the log-log plot leading to an exponent $1/\alpha$ slowly varying between $3/2$ and $2$. Accordingly the spectrum develops larger tails and there is a redistribution of spectral weight towards larger eigenvalues. The situation is clearly different from the crystal case were  $1/\alpha = 1$. \
\item {\it Mode extension}: The closest the system is to the jamming transition, the more coherent and spatially organized the softest mode is. This is visually clear on Fig.~\ref{modes_dif_phi}, which provides an example of the softest mode for each packing fraction. 
\end{itemize}
Two quantitative results support this last assertion. First, the participation ratio for the largest modes increases from $P(1)=0.2$ to $P(1)=0.4$ when approaching $\phi_J$. Second, spatial correlations within the modes increase. This is measured by computing the following correlation function:
\be
C_m(r) = \left\langle (\vec u_m^i - \langle \vec u_m\rangle) . (\vec u_m^j - \langle \vec u_m\rangle)\right\rangle_{i,j/ d_{i,j} = r},
\ee
where the average is computed over all pairs of particles separated by $r$. These spatial correlators are plotted for $m=1$ and $m=10$ in the insets of Fig.~\ref{modes_dif_phi}. Clearly, the correlation extends on a longer distance when the system is closer to $\phi_J$. An interesting feature is that $C_m(r)$ becomes negative for $r\approx 10$, indicating some anti-correlation, which we attribute to the vortices pattern observed in the modes. Also, the correlation is much weaker for $m=10$ than for $m=1$. This effect is further characterized in the main plot of the same figure, where we plot $I_m= \sum_{r<5} C_m(r)$ versus $m$ for the four packing fractions. Not only the modes have a structure on a larger scale closer to $\phi_J$, but also more of them are structured.

\subsection{Soft modes and dynamics in a metastable state}

\begin{figure}[t!]
\begin{center}
\rotatebox{-90}{\resizebox{!}{8cm}{\includegraphics{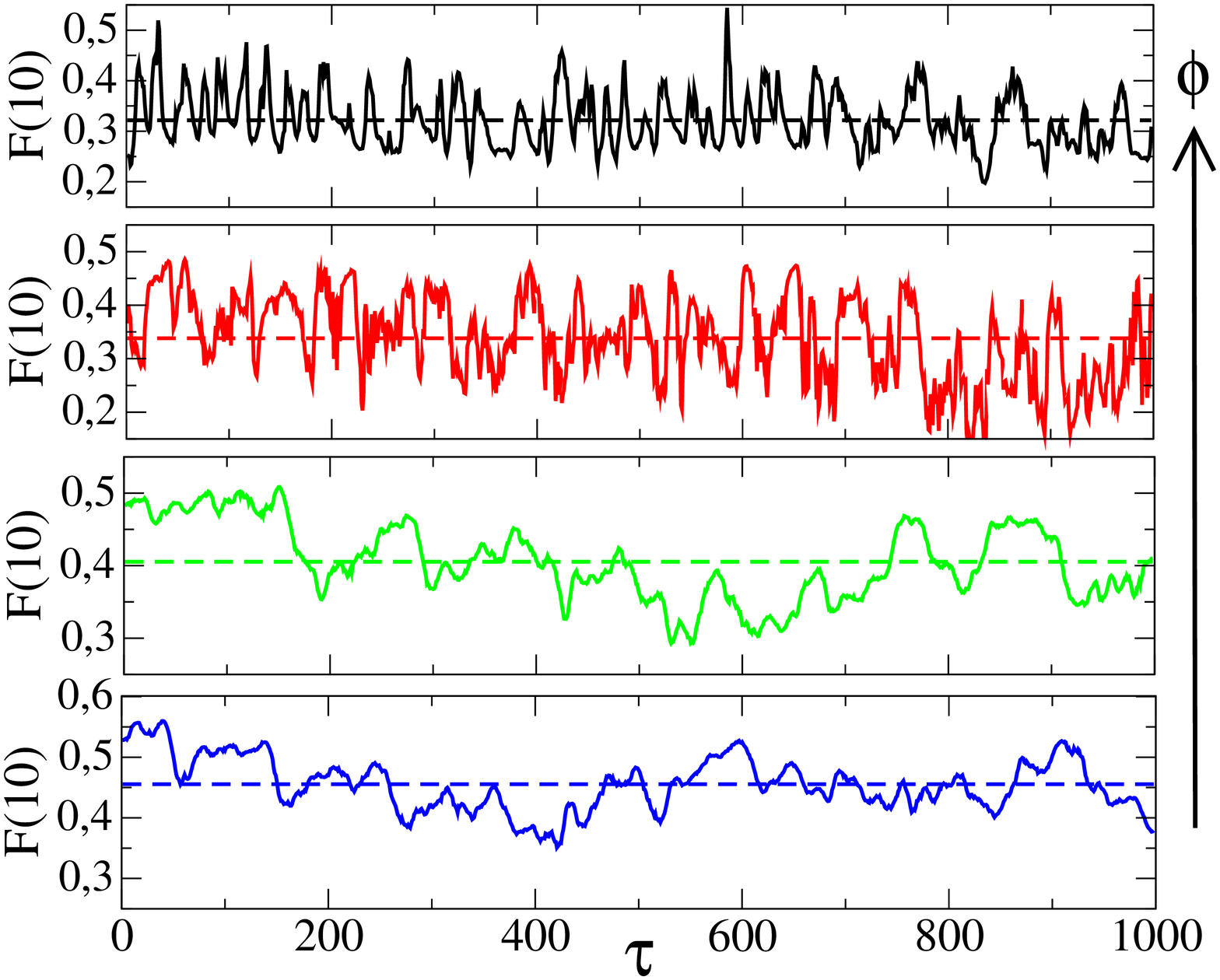}}}
\rotatebox{-90}{\resizebox{!}{8cm}{\includegraphics{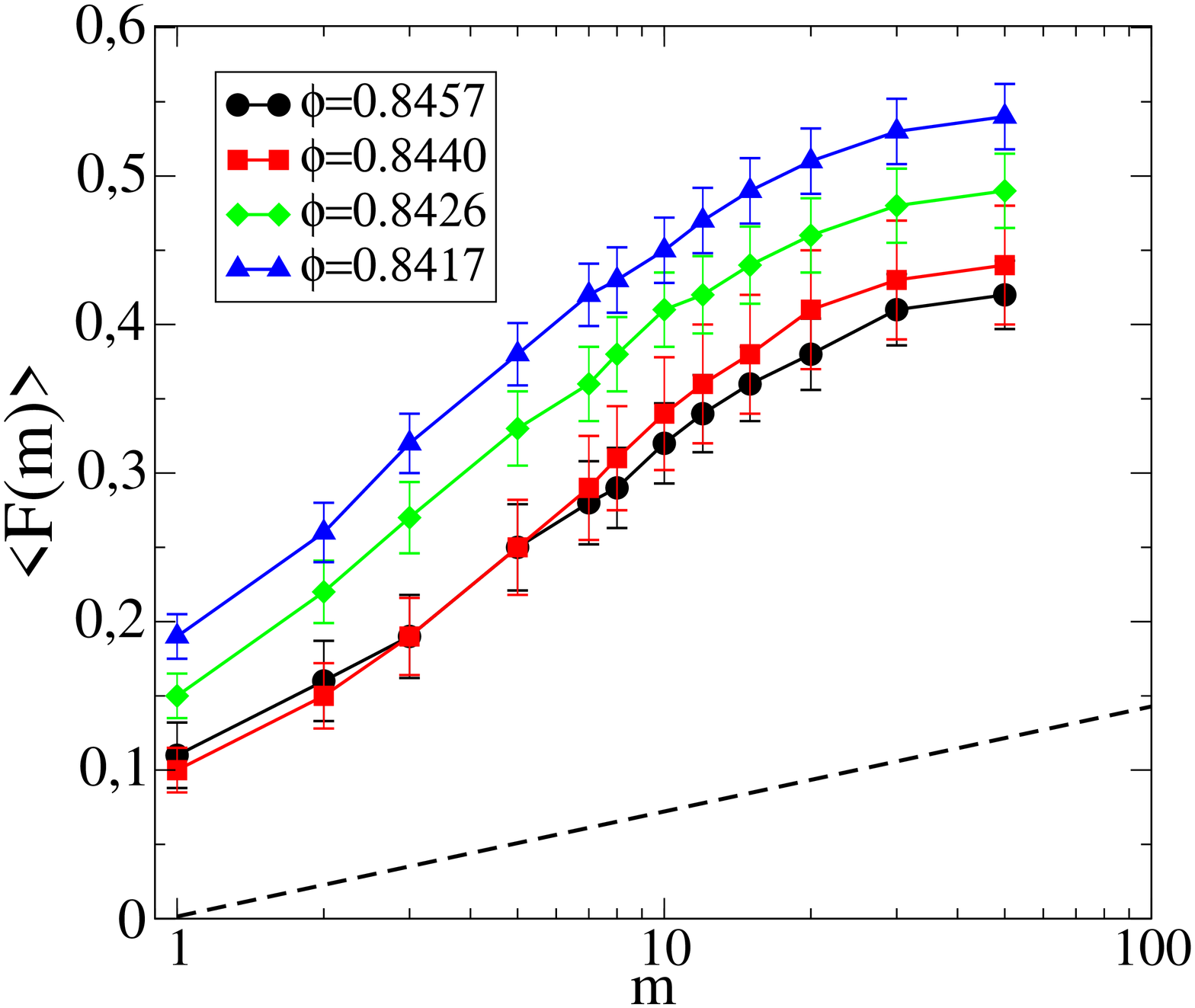}}}
\caption{Projection of the dynamics restricted to the rigid subsets. {\bf Top}: Fraction of the dynamics projected on the $10$ most significant mode: $\langle F(10) \rangle$ vs. $\tau$ as defined in the text. The arrow on the right indicates the direction of increasing packing fraction $\phi \in [0.8417, 0.8426, 0.8440, 0.8457]$. {\bf Bottom}: Average value of $\langle F(m) \rangle$ vs. $m$ for different $\phi$. The error bars correspond to the standard deviations of $F(m)$ and the dashed line is the prediction of Eq. (\ref{fm}) for a basis composed of random modes.}
\label{modos_dinamica}
\end{center}
\end{figure}

We now turn to the relation between the modes $|\lambda_m\rangle$ and the dynamics $|r(t)\rangle=\{\vec{r}_i(t)\}$. We first concentrate on the dynamics restricted to the rigid subsets and consider the projection of the real dynamics on the modes computed in an observation window preceding the dynamics by a lag time $\tau$. More precisely, let $[t,t+T]$ be the time window where the basis of eigenmodes $\{|\lambda_m \rangle\}$ is computed. The dynamical evolution $|\delta r(\tau)\rangle = |r(t+T+\tau)\rangle - |r(t+\tau)\rangle $ is then projected on the modes $m$ and the corresponding component is rescaled by the amplitude of the dynamics:
\be
c_m \equiv \frac{\langle\lambda_m|\delta r(\tau)\rangle}{\langle\delta r(\tau)|\delta r(\tau)\rangle}.
\ee
The components $c_m$ satisfy $\sum_m (c_m)^2=1$ since the eigenvectors form a complete basis.          
We sort the $c_m$  in decreasing order ${c_k^0}={c_1>c_2...>c_{2N}}$ and, following~\cite{Brito_JCP2009}, define:
\be                       
F(m) = \sum_{k=1}^m c_k^2.
\label{fm}
\ee                       
$F(m)$ measures the fraction of the dynamics ``explained'' by the $m$ most contributing modes.  
Here, in the light of the previous section, we have chosen to consider the $10$ first modes, $T=100$,  
$\tau$ varies from 1 to 1,000 cycles and we average $F(10)$ on two initial times $t$ as well as over all the rigid subsets. Figure~\ref{modos_dinamica}-top displays $F(10)$ for the four packing fractions. Three key aspects emerge:
\begin{itemize}
\item For all packing fractions $F(10)$ fluctuates around a large constant value. This shows that even for large $\tau$ the dynamics is well described by the $10$ most significant modes of ${\bf C}_p$ as long as the system remains in a metastable state. From this perspective, the modes defined by ${\bf C}_p$ for the rigid subsets give a faithful representation of the dynamics and can be indeed considered as effective vibrational modes. 
\item Interestingly, the average value $\langle F(10) \rangle_{\tau}$ increases, beyond error bars, when $\phi$ increases towards $\phi_J$. This indicates that the $10$ first modes concentrate a more important part of the dynamics as $\phi \to \phi_J$, in agreement with the idea that the dynamics becomes more collective, or structured. This is further demonstrated on Figure~\ref{modos_dinamica}-bottom where $\langle F(m) \rangle$ is plotted versus $m$ for the fifty largest modes. The closer to $\phi_J$, the larger is $\langle F(m) \rangle$.
\item The fluctuations of $F(10)$ are clearly more correlated in time when $\phi$ decreases, revealing that the modes have a larger characteristic time closer to $\phi_J$.
\end{itemize}

\begin{figure}[t!]
\begin{center}
\rotatebox{-90}{\resizebox{!}{8cm}{\includegraphics{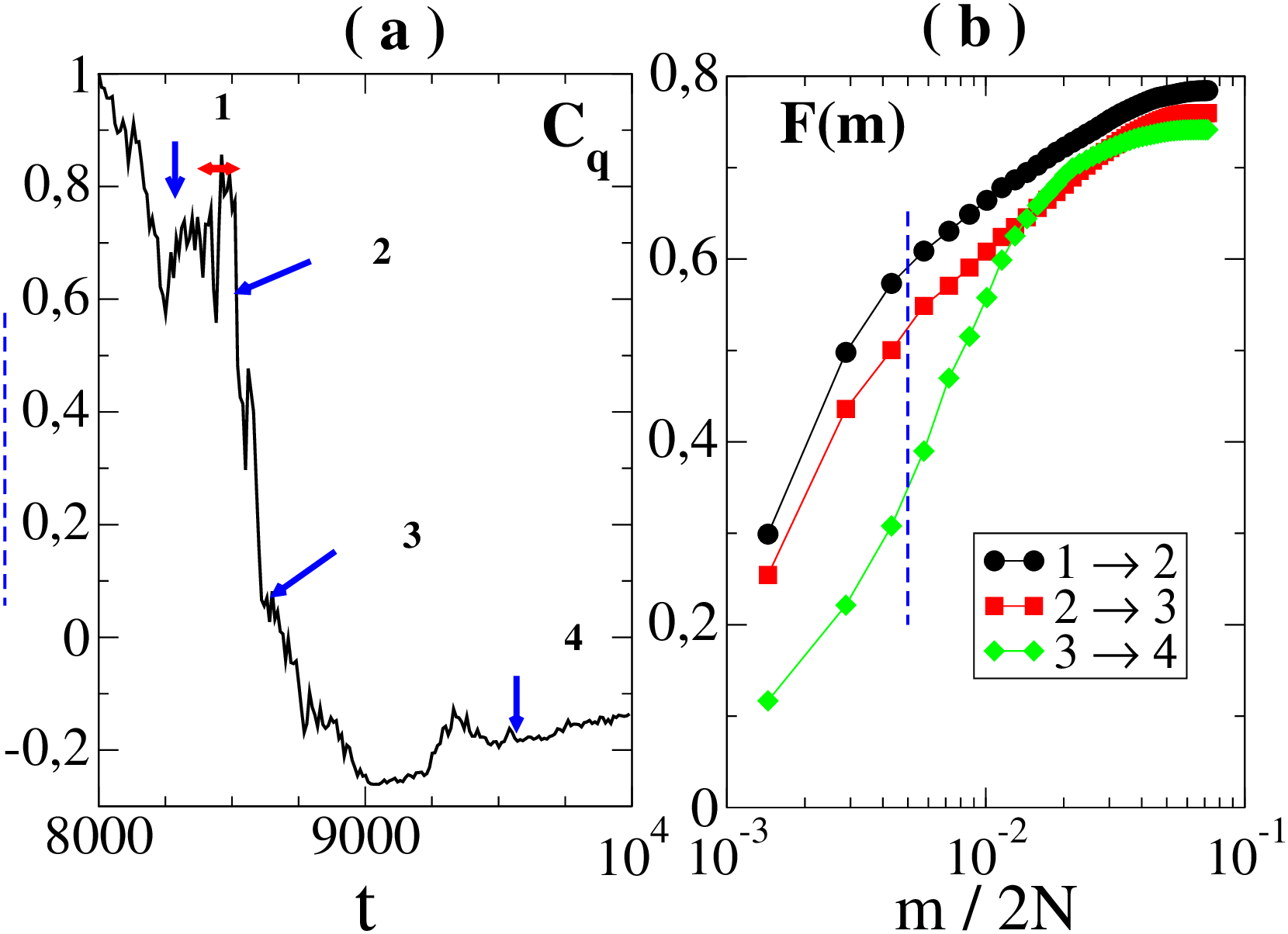}}}
\rotatebox{-90}{\resizebox{!}{8cm}{\includegraphics{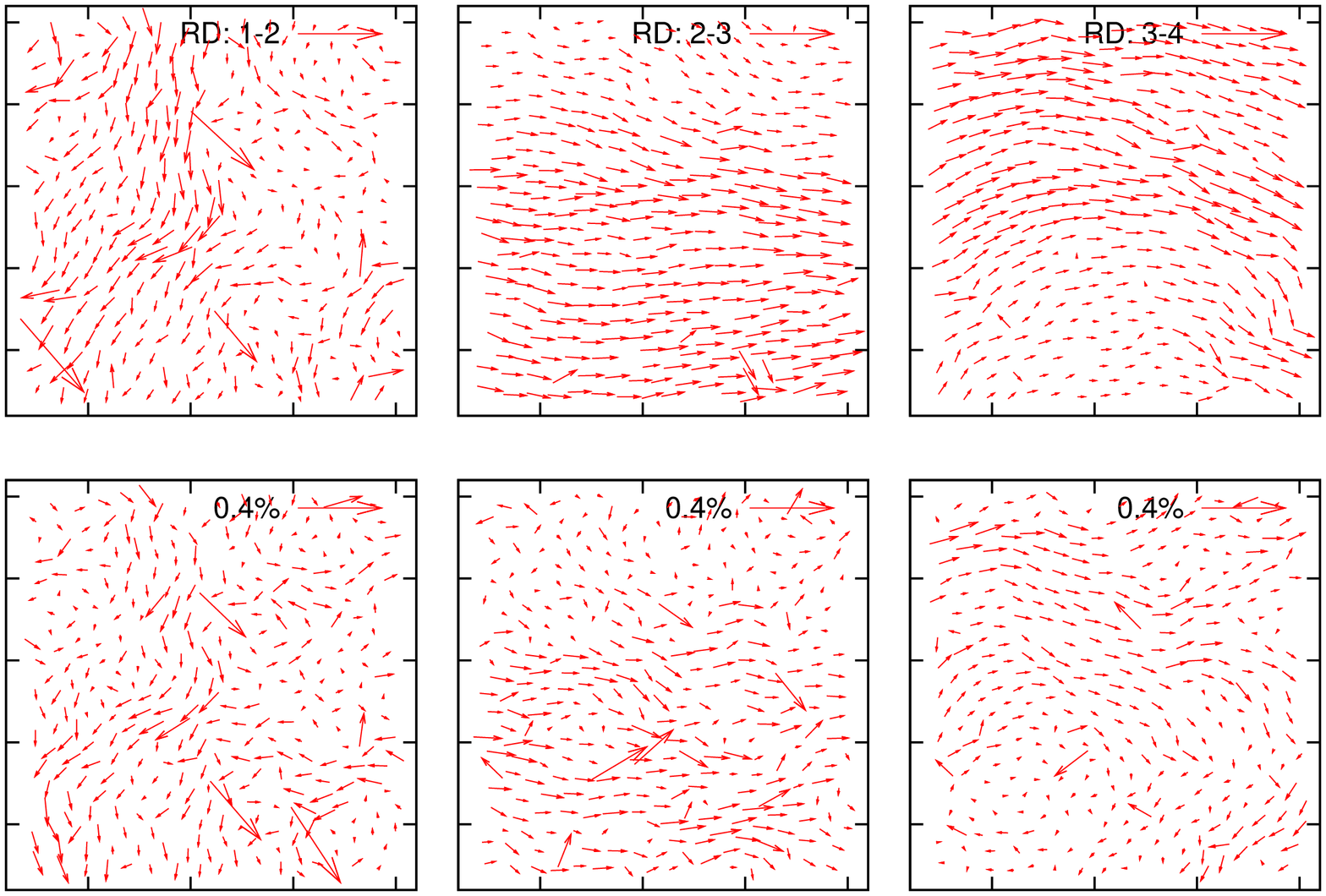}}}    
\end{center}
\caption{Analysis of a relaxation event within a sub-region, following a quiet period  $\phi=0.8417$. 
Above: {\bf (a)} $C_q(t)$ indicates a sudden collective motion of the particles, the so-called ``crack". The four arrows enumerated from 1 to 4 indicate the bounds of the time intervals, where the real dynamics is considered. The quiet period during which the modes are computed is indicated by the straight horizontal (red) line just preceding the crack. {\bf (b)} $F(m)$ vs. $m/2N$ for the three dynamics defined above; the vertical blue dotted line indicates $m=3$. {\bf (c)}: On top of each other the real dynamics during the three periods of time and its recomposition with the 3 most significant modes}
\label{crack_data12}
\end{figure}

\subsection{Soft modes and cracks: a preliminary analysis}

The observations above suggest that when approaching $\phi_J$ a smaller and smaller amount of modes concentrate the dynamics for longer and longer times : the system ``rattles'' around its metastable state along more and more preferential and softer directions in phase space. 
However, for all the methodology issues seen above, and because of the lack of theoretical grounds for frictional systems, determining whether such directions determine the way the system locally cracks (as seen in other systems~\cite{Brito_JCP2009, ISI:000259686400015}) is an extremely challenging issue in the present system. Here we provide a first attempt to answer this question, which obviously deserves further analysis. 

We identify in the self-density correlation function $C_q(t)$ a sudden drop preceded by a quiet period (see Fig.~\ref{crack_data12}-a). We compute the covariance matrix of the positions restricted to this rigid subset before the crack $T_<=[t_1,t_2]$. We then consider two further time intervals related to the crack: one during the crack $T_{=}=[t_2,t_3]$ and one after the crack $T_>=[t_3,t_4]$. We observe the real dynamics during these intervals and project the dynamics onto the first $m$ modes determined before the crack, defining $F_<(m),F_=(m)$ and $F_>(m)$. One observes on figure~\ref{crack_data12}-(b) that $F_=(m)$ and $F_<(m)$ share a similar behaviour as a function of $m$, different from the one of $F_>(m)$: both $F_<(m)$ and $F_=(m)$ increase sharply at small $m$ whereas $F_>(m)$ only increases for larger $m$. In all cases $70\%$ of the dynamics is explained by the first ten modes, but only $35\%$ of the dynamics taking place after the crack projects on the first three modes, whereas this fraction reaches $60\%$ for the dynamics taking place before or during the crack. A visual transcription of these numbers is provided by figure~\ref{crack_data12}-(c). 

These observations suggest that (i) the crack considered here is really a micro-event in the sense that the dynamics after the crack still projects on a small amount of modes (here of the order of ten), (ii) at least for such a micro-crack, there is a selection of directions in phase space along which the cracks occurs.

\section{Discussion and conclusion}

This paper is a first attempt to describe the dynamics of granular media in the jammed, glassy region in terms of ``modes'', by applying a Principal Component Analysis (PCA) to the covariance matrix of the position of individual grains. This is perfectly justified, and gives sensible results, in a regime of time/densities such that the average position of the particles is approximately constant, that is, varies less than the typical fluctuations themselves, otherwise both the reference configuration and covariance matrix itself evolve with time. The time scale over which the reference configuration can be considered as stable also depends on the system size, since in an infinite system some rearrangement takes place somewhere in the system at each instant of time. 

For small enough times/system sizes, or at high enough packing fractions, this stability criterion is approximately fulfilled and the spectral properties of the covariance matrix reveals large, collective fluctuation modes that cannot be explained by a Random Matrix benchmark where these correlations are discarded. The existence of these collective modes is expected from the results of \cite{fred1} that established the existence of dynamical correlations, which diverge as the system reaches its rigidity transition $\phi_J$.  The analysis in terms of eigenmodes provided here confirm that the slow, large scale dynamic structures appear when $\phi \to \phi_J^+$, that explain a substantial fraction of the dynamics.

We then attempted to find some link between the softest modes of the covariance matrix during a certain ``quiet'' time interval and  the spatial structure of the rearrangement event that ends this quiet period. In order to do so, we first tried to identify well-defined ``cracks'' that would lend themselves to such an analysis. This proves to be exquisitely difficult: the rearrangements are made of micro-cracks of all amplitudes, that span larger and larger regions of the system as $\phi \to \phi_J$ and that are at the origin of the superdiffusive, L\'evy flight character of the motion found in \cite{JP-Levyflights}. In spite of this difficulty, we have succeeded in identifying some ``rigid subsets'' where well characterized cracks appear. The motion during these cracks is indeed well explained by the soft modes of the dynamics before the crack. However, a more systematic analysis should be undertaken because we do not know at this stage whether the identification of these rigid subsets induces a strong selection bias on the nature of the cracks themselves. In the hypothesis where for a majority of cracks we could even not define the precursor modes then these soft modes would not be relevant to understand the dynamical evolution of the  system. We believe that this is increasingly the case as one approaches the rigidity transition, where self-similar micro-cracks of all scales become overwhelming. In that eventuality, the analysis in terms of mode would only be useful to characterize the rigidity of amorphous granular systems, for dense enough packings where the rigid subsets remain dominant. In all cases, we believe that the methodology presented here will motivate and buttress further work in that direction.  

\vspace{-5mm}
\acknowledgements
We would like to thank Silke Henkes and Wim van Saarloos for highly valuable discussions. O. Dauchot is grateful to KNAW for his visiting position in Leiden. During the very last stages of the present work, he became aware of a similar work being performed in the case of NIPA colloidal particles in Andrea Liu's group, in Pennsylvania University. This was the opportunity to perform with S. Henkes and W. van Saarloos the same kind of comparison analysis of the experimental spectra to those obtained for a simulated Random Matrix model. We would like to thank the members of this group for having shared with us some of their data and discuss the issues relative to their analysis. C. Brito stays in Saclay, respectively in Leiden, were supported by a grant from the RTRA ``Triangle de la Physique", respectively from FOM.

\bibliography{modes_references}

\end{document}